\documentclass[dvips]{article}
\usepackage{graphicx}
\usepackage{amssymb}
\usepackage{graphicx}
\usepackage{dcolumn}
\usepackage{amsmath}
\usepackage{lscape}
\usepackage{longtable}
\usepackage{subfigure}
\usepackage{color}

\begin{document}
\newcommand{\bd}{\begin{document}}
\newcommand{\ed}{\end{document}}
\newcommand{\bc}{\begin{center}}
\newcommand{\ec}{\end{center}}
\newcommand{\bfr}{\begin{flushright}}
\newcommand{\efr}{\end{flushright}}
\newcommand{\lt}{\left}
\newcommand{\rt}{\right}
\newcommand{\vs}{\vspace}
\newcommand{\hs}{\hspace}
\newcommand{\beq}{\begin{equation}}
\newcommand{\eeq}{\end{equation}}
\newcommand{\lb}{\linebreak}
\newcommand{\pb}{\pagebreak}
\newcommand{\mb}{\makebox}
\newcommand{\fb}{\framebox}
\newcommand{\mc}{\multicolumn}
\newcommand{\ben}{\begin{enumerate}}
\newcommand{\een}{\end{enumerate}}
\newcommand{\bit}{\begin{itemize}}
\newcommand{\eit}{\end{itemize}}
\newcommand{\ol}{\overline}
\newcommand{\un}{\underline}
\newcommand{\lefq}{\lefteqn}
\newcommand{\ba}{\begin{array}}
\newcommand{\ea}{\end{array}}
\newcommand{\beqa}{\begin{eqnarray}}
\newcommand{\eeqa}{\end{eqnarray}}
\newcommand{\beqas}{\begin{eqnarray*}}
\newcommand{\eeqas}{\end{eqnarray*}}
\newcommand{\bfg}{\begin{figure}}
\newcommand{\efg}{\end{figure}}
\newcommand{\bds}{\begin{displaymath}}
\newcommand{\eds}{\end{displaymath}}
\newcommand{\btb}{\begin{tabbing}}
\newcommand{\etb}{\end{tabbing}}
\bc {
\textbf{\huge  The  massless Dirac-Weyl equation with deformed extended complex potentials} } \ec

\vs{1cm}

\bc
{\it \"Ozlem Ye\c{s}ilta\c{s}$^{*}${\footnote {e-mail : yesiltas@gazi.edu.tr, bengudemircioglu@yahoo.com} and Beng\"{u}  \c{C}a\~{g}atay$^{\dag}$   \\
$^{*}$Department of Physics, Faculty of Science, Gazi University,
06500 Ankara, Turkey\\$^{\dag}$ Turkish Atomic Energy Authority, \\ Sarayk\"{o}y Nuclear Research and Training Centre, 06983  Ankara, Turkey\\
\vspace{.16cm}

}} \ec \vs{1cm}
\begin{abstract}
\noindent Basically (2 + 1) dimensional Dirac equation with  real  deformed  Lorentz scalar
potential is  investigated in this study.  The position dependent Fermi velocity function transforms Dirac Hamiltonian into a
 Klein-Gordon-like effective Hamiltonian system. The complex Hamiltonian and its real energy spectrum and eigenvectors are  obtained analytically.
Moreover, the Lie algebraic analysis is also performed.
\end{abstract}
\noindent {\bf keyword:}   Dirac equation, massless fermion, analytical solutions, Lie algebras \\


\section{Introduction}
In physics, spin 1/2 massless particles  are represented by a two component Weyl spinor. In the motivation of quantum field theory and condensed matter physics, the quasi particles which are massless are of  interest because Dirac and Weyl materials are related to the ones with  low-energy quasiparticle excitations which necessity the Dirac and Weyl equations for the massless fermions moving through the graphene surface. A new invention which is discovery of graphene in 2004 has attracted an enormous attention of the physicists because of its extraordinary properties \cite{novo}. Graphene has got a honeycomb structure with a single layer of carbon atoms is a two dimensional material. Then, the study of Dirac-Weyl particles in magnetic fields has received much attention  for confining the charges \cite{pinaki}, \cite{roy}, \cite{kuru}, \cite{midya}.  Moreover, the Dirac theory in low dimensions is studied for the different potentials both numerically \cite{h1} and  theoretically \cite{h2}, \cite{h3}. Also, graphene with the zero energy states is searched and scalar potentials are generated within the exactly solvable models \cite{axel}. On the other hand, pseudo-particles have a Fermi velocity $~10^{6} m/s$ and they can be replaced with the position dependent one as given in \cite{lima}, \cite{pan}, \cite{1} and \cite{yes}.  And, rationally extended potentials have been attracted too much attention in the literature \cite{axel1}, \cite{axel3}, \cite{yad}, \cite{mid}.  The Dirac Hamiltonian becomes as non-Hermitian with position dependent Fermi velocity and in this study we will investigate the Dirac Hamiltonian with a complex deformed $\mathcal{PT}$ symmetric \cite{pt} potential which is also rationally extended. Using the Lie algebraic analysis, we have obtained the $so(2,1)$ operators for the extended complex deformed $\mathcal{PT}$ symmetric potential and arrived at the spectrum.
\section{Dirac Equation }
The effective Hamiltonian has a form $H=\pm \hbar v_{F}\vec{\sigma}\cdot (-i \hbar \vec{\nabla})$ for graphene where an electron moves with an effective Fermi velocity  $v_{F} = c/300 m/s $,  c is the velocity of light and behaves as a massless quasi-particle. On the other hand one can consider the Hamiltonian including position dependent Fermi velocities instead of the constant velocity. Now, the Dirac-Weyl equation for the massless particles under an external magnetic field is  given   below
\begin{equation}\label{1}
  \textbf{H}=\sqrt{\mathcal{V}_{F}(x)}[\overrightarrow{\sigma}\cdot (\overrightarrow{p}\sqrt{\mathcal{V}_{F}(x)}+i\frac{e}{c}\overrightarrow{A})]
\end{equation}
where $\mathcal{V}_{F}(x)$ is the position dependent Fermi velocity \cite{1}, $\overrightarrow{\sigma}=[\sigma_{x},\sigma_{y}]$ are the Pauli matrices, $\overrightarrow{p}$ is the momentum operator and we give the vector potential $\overrightarrow{A}$ as $\vec{A}=[0,A_{y}(x),0]$ which is normal to the surface. The Fermi velocity is measured for different graphene layers and the changing Fermi velocity cases are discussed in \cite{lima}, \cite{F}. The eigenvalue equation can be shown as $ \textbf{H} \Psi(x,y)=E \Psi(x,y)$. Here, the two component Dirac wave-function is given by $\Psi(x,y)=e^{ik_{y}y}\left(
                         \begin{array}{c}
                           \psi_{A}(x) \\
                           \psi_{B}(x) \\
                         \end{array}
                       \right)$. And the eigenvalue equation turns into a couple of second order Sturm-Liouville type  equations which are represented in a single equation below
\begin{equation}\label{5}
  -\mathcal{V}^{2}_{F}(x)\psi''_{A,B}(x)-2\mathcal{V}_{F}(x) \mathcal{V}'_{F}(x)\psi'_{A,B}(x)+U_{eff, A,B}\psi_{A,B}(x) = E^{2}\psi_{A,B}(x)
  \end{equation}
where the effective potentials can be written as
\begin{eqnarray}\label{6}
  U_{eff,A}(x) &=& -(e\bar{A}(x)-ik\mathcal{V}_{F}(x))^{2}+i\mathcal{V}_{F}(x)(e\bar{A}'(x)-ik\mathcal{V}'_{F}(x))-\frac{1}{4}\mathcal{V}'^{2}_{F}(x)-
  \frac{1}{2}\mathcal{V}_{F}(x)\mathcal{V}''_{F}(x) \\
  U_{eff,B}(x) &=&  -(e\bar{A}(x)-ik\mathcal{V}_{F}(x))^{2}-i\mathcal{V}_{F}(x)(e\bar{A}'(x)-ik\mathcal{V}'_{F}(x))-\frac{1}{4}\mathcal{V}'^{2}_{F}(x)-
  \frac{1}{2}\mathcal{V}_{F}(x)\mathcal{V}''_{F}(x).
\end{eqnarray}
We can apply a subsequent transformation to (\ref{5}) which is $\psi_{A,B}(x)=\frac{\phi_{A,B}(x)}{\sqrt{\mathcal{V}_{F}(x)}}$, $z=\int\frac{dx}{\mathcal{V}_{F}(x)}$ and we get
\begin{equation}\label{7}
  -\phi_{A,B}''(z)+U_{A,B}(z)\phi_{A,B}(z)=E^{2}\phi_{A,B}(z),
\end{equation}
where
\begin{eqnarray}\label{8}
  U_A(z) &=& -e^{2}A(z)^{2}+2iekA(z)\mathcal{V}_{F}(z)+k^{2}\mathcal{V}^{2}_{F}+ieA'(z)+k\mathcal{V}'_{F}(z)= W_{1}(z)^{2}+W_{1}'(z)\\
 U_B(z) &=& -e^{2}A(z)^{2}+2iekA(z)\mathcal{V}_{F}(z)+k^{2}\mathcal{V}^{2}_{F}-ieA'(z)-k\mathcal{V}'_{F}(z)=W_{1}(z)^{2}-W_{1}'(z), \label{80}
\end{eqnarray}
where the superpotential is $W_{1}(z)=i(eA(z)-ik\mathcal{V}_{F}(z))$. Now we will consider two models which are known as deformed Scarf and P\"{o}schl-Teller potentials. Here, deformed trigonometric functions and their relations can be shown by $ \cosh_{q} x=\frac{e^{x}+qe^{-x}}{2},~~~~\sinh_{q} x=\frac{e^{x}-qe^{-x}}{2},~~~~\cosh^{2}_{q}x-\sinh_{q}^{2} x=q,~~~~\tanh_{q} x=\frac{\sinh_{q} x}{\cosh_{q} x}$ and one can look at the details given in \cite{arai}, \cite{cari} and for other works in \cite{h4}, \cite{h5}.
\subsection{$\mathcal{PT}$ symmetric hyperbolic deformed Scarf II potential } Let's choose $\mathcal{V}_{F}(z)=K_1 \tanh_{q}(z), A(z)=K_2\ sech_{q}z$, $K_1, K_2$ are real  parameters. Using $W^{2}_{1}(z) \pm W_{1}'(z)$, (\ref{8}) and (\ref{80}) become
\begin{eqnarray}\label{9}
  U_A(z) &=&i e K_2 (2k K_1-1)\sec h_{q}z \tanh_{q}z-(e^{2}K^{2}_{2}-kK_{1})\sec h^{2}_{q}z+K^{2}_{1}k^{2}\tanh^{2}_{q} z\\
 U_B(z) &=& i e K_2 (2k K_1+1)\sec h_{q}z \tanh_{q}z-(e^{2}K^{2}_{2}+k^{2}K_{1})\sec h^{2}_{q}z+K^{2}_{1}k^{2}\tanh^{2}_{q} z. \label{90}
\end{eqnarray}
And the solutions for $U_B(z)$ are well-known \cite{lev}. It is known that the Hamiltonians $H^{(1)}_{-}=\hat{A}_{1}^{\dag}\hat{A}_{1}=-\frac{d^{2}}{dz^{2}}+V^{(1)}_{-}(z)$, $H^{(1)}_{+}=\hat{A}_{1} \hat{A}_{1}^{\dag}=-\frac{d^{2}}{dz^{2}}+V^{(1)}_{+}(z)$ where $\hat{A}_{1}=\frac{d}{dz}+W_{1}(z)$ and the potential $U_{B}(z)=V^{(1)}_{-}(z)$ is the element of $H^{(1)}_{-}$. Now we will search for more general complex potentials, then,  $W_{2}(z)$ super-potential in terms of the unknown parameters as
\begin{equation}\label{10}
  W_{2}(z)=C_1 \tanh_{q} z+ i C_2 \sec h_{q} z + \frac{C_3 \cosh_{q} z}{r_1+r_2 \sinh_{q} z}.
\end{equation}
Then using supersymmetric quantum mechanics, we obtain $V^{(2)}_{-}(z)=W_{2}^{2}(z)-W_{2}'(z)$ as
\begin{equation}\label{11}
  V^{(2)}_{-}(z)=i(2C_1+1)C_2\sec h_{q} z \tanh_{q} z-(C_1+C^{2}_{2}+qC^{2}_{1})\sec h_{q}^{2}z+C^{2}_{1}+C_3\frac{K_1+K_2 \cosh_{q} 2z+ K_3 \sinh_{q} z}{2(r_1+r_2 \sinh_{q} z)^{2}}
\end{equation}
where the constants on the nominator of the rational term are $K_1=C_3+4iC_2r_1+2r_2-2C_1 r_2$, $K_2=C_3+2C_1 r_2$ and $K_3=-2r_1+4C_1 r_1+4i C_2 r_2$ and because we want these constants have to vanish, then, we can obtain $C_3=-i(\mp r_1+2C_2 r_1), C_1=\frac{1}{2}(1\mp 2C_2), r_2 \rightarrow \mp i r_1$. We note that $C_1$ and $C_2$ are real parameters. Then we have a couple of isospectral partner potentials which are given by
\begin{equation}\label{12}
  V^{(2)}_{-}(z) = 2iC_{2}(1\mp C_2)\sec h_{q} z \tanh_{q} z+\frac{1}{4}(1\mp 2C_2)^{2}-\frac{1}{2}(1+2C^{2}_{2} +\pm2 C_2+q\frac{(1 \mp 2C_2)^{2}}{2} )\sec h_{q}^{2}z,
  \end{equation}
  \begin{equation}\label{13}
  \begin{split}
    V^{(2)}_{+}(z) =  \mp 2i C^{2}_{2} \sec h_{q} z \tanh_{q} z+\frac{1}{4}(1 \mp 2C^{2}_{2})+\frac{1}{2}(1+2C_2(-C_2 \mp 1)+\frac{q}{2}(1 \mp C_2)^{2})\sec h_{q}^{2} z+ \\ \frac{(2C_2 \mp 1)(\mp 1+2i \sinh_{q}z)}{(-i \mp \sinh_{q}z )^{2}}
    \end{split}
  \end{equation}
Here, it is given as $-\infty < z< +\infty$, then, (\ref{12}) and (\ref{13}) are isospectral potentials except the ground state energy. Comparing (\ref{12}) and (\ref{90}), we can express the superpotential parameters in terms of our system parameters, then the  real energy eigenvalues can be given by \cite{yad}
\begin{equation}\label{14}
  E^{(2)}_{(-),n}=E^{(1)}_{(-),n}= \pm \sqrt{\frac{1}{4}(1 \mp 2C_2)^{2}-(\lambda_{1}-n)^{2}}, ~~~~E^{(2)}_{(+),n}=E^{(2)}_{(-),n+1}
\end{equation}
where $\lambda_{1}=\frac{1}{2}(-1-2\lambda_{2} \mp \sqrt{1+4k^{2}K_1+4eK_2+8ekK_1 K_2+4e^{2}K^{2}_{2}})$, $ \lambda_{2}$ is a constant. We shall compare (\ref{12}) and (\ref{90}) and obtain
\begin{equation}\label{15}
K_1 = \frac{e(1+e)-2e^{3}-e^{4}-k+eq(e-1)+\frac{1}{2}\sqrt{-8(e^{2}-1)(-1+2e^{2}-k)(e^{2}-q)+4(2e^{3}+k+e(q-1)-q)^{2}}}{2(e^{2}-1)k(e^{2}-q)}
 \end{equation}
\begin{equation}\label{16}
   C_2 = \frac{e^{2}(1-2e^{2}+e(q-k)-e^{2}q+\frac{e}{2}\sqrt{-8(e^{2}-1)(-1+2e^{2}-k)(e^{2}-q)+4(2e^{3}+k+e(q-1)-q)^{2}}}{2(e^{2}-1)k(e^{2}-q)},
\end{equation}
where we choose $K_2=1+C_2$ for the sake of simplicity in our solutions. One can use (\ref{16}) in (\ref{14}) to express the energy spectrum in terms of the initial potential parameters in the Dirac equation. Using the well-known formula for the ground-state wavefunction $\psi^{(1)}_{(-),0}(z)=\exp(\int^{z} W_{1}(t) dt)$, we get
\begin{equation}\label{gs}
  \psi^{(1)}_{(-),0}(z) = \exp[2ieK_{2}\arctan (\tanh_{q}(\frac{z}{2}))] (\sec h_{q} z)^{-kK_{1}}.
\end{equation}
Then, we can use the ground-state solution in order to find the general solutions, $\psi^{(1)}_{(-),n}(z)=\psi^{(1)}_{(-),0}(z) P_{n}(z)$ where $P_{n}(z)$ is an unknown polynomial in the Hamiltonian equation $H^{(1)}_{-} \psi_{1}=E^{2}\psi_{1}$, hence we can obtain a solution. Or, because we have already compared our system with the spectrum results given in \cite{yad}, we can obtain the exact solutions for our deformed system, therefore,
the spinor solutions $\psi_{B}(z)$ can be also written as \cite{yad}
\begin{equation}\label{17}
  \psi_{B, n}(z)=\psi^{(1)}_{(-),n}(z)=\psi^{(2)}_{(-),n}(z)=N_{n} (\sec h_{q} z)^{\lambda_{1}}\exp(-i\lambda_{2} \arctan(\sin h_{q} z ))P^{(\lambda_{2}-\lambda_{1}-1/2,-\lambda_{2}-\lambda_{1}-1/2)}_{n}(i \sinh_{q} z),
\end{equation}
where $N_{n}$ is the normalization constant, $P^{(\alpha, \beta)}_{n}(z)$ are the Jacobi polynomials. We also remind that the constants $\lambda_{1}$ and $\lambda_{2}$ are found before in terms of the initial system parameters. And, for the system (\ref{13}), $\psi_{+}(z)$ can be found as
\begin{equation}\label{18}
  \psi^{(2)}_{(+),n+1}(z)=\frac{1}{\sqrt{E^{(2)}_{(-),n}}}\hat{A}_{2}\psi^{(2)}_{(-),n}(z)=N'\frac{(\sec h_{q} z)^{\lambda_{1}}\exp(-i\lambda_{2} \arctan(\sin h_{q} z)}{-i \mp \sinh_{q}z }P^{(\lambda_{1}, \lambda_{2})}_{n+1}(i \sinh_{q} z).
  \end{equation}
\section{Lie algebras}
The operators of $iso(2,1)$ Lie algebra are given by
\begin{equation}\label{30}
  J_{\pm}= ie^{\pm i\phi}(\pm \frac{\partial}{\partial z}+((-i\frac{\partial}{\partial \phi}\pm \frac{1}{2})F(z)-G(z))
\end{equation}
and
\begin{equation}\label{31}
  J_3=-i\frac{\partial}{\partial \phi}.
\end{equation}
They obey the commutation rules given below as
\begin{equation}\label{32}
   [J_{+}, J_{-}]=-2J_{3},~~~~[J_{3}, J_{\pm}]=\pm J_{\pm}.
\end{equation}
In \cite{yad}, the operators of the $so(2,1)$ algebra is extended as
\begin{equation}\label{33}
  J_{\pm}= e^{\pm i\phi}(\pm \frac{\partial}{\partial z}-((-i\frac{\partial}{\partial \phi}\pm \frac{1}{2})F(z)-G(z))+U(z,-i\frac{\partial }{\partial \phi} \pm 1/2)).
\end{equation}
Now,  $U(z,-i\frac{\partial}{\partial \phi} \pm 1/2)$ is the extra operator which is used in \cite{yad}. The term $U(z,-i\frac{\partial}{\partial \phi} \pm 1/2)$ plays a role that  it helps to construct the algebra for the rationally
extended potentials. Here, we discuss the extended hyperbolic complex Scarf II potential within the $so(2,1)$ algebra. If $J_{\pm}$ and $J_{3}$ are used in (\ref{32}), then, one can get
\begin{equation}\label{34}
  F'(z)-F(z)^{2}=1,~~~~G'(z)-F(z)G(z)=0.
\end{equation}
The constraints in (\ref{32}) also lead to \cite{yad}
\begin{equation}\label{35}
\begin{split}
  U_{1}(z)^{2}-\frac{d}{dz}U_{1}(z)+2U_{1}(z)(F(z)(\mu+\frac{1}{2})-G(z))-\\
  (U_{2}(z)^{2}+\frac{d}{dz}U_{2}(z)+2U_{2}(z)(F(z)\mu_1-G(z)))=0.
\end{split}
\end{equation}
In this study we will use $U(z,-i\frac{\partial}{\partial \phi}-1/2=U_{1}(z)$ and $U(z,-i\frac{\partial}{\partial \phi}+1/2=U_{2}(z)$.  And the Hamiltonian  $H^{(2)}_{-}=-\frac{d^{2}}{dz^{2}}+V^{(2)}_{-}(z)$ can be given in terms of the Casimir operator
$J^{2}=J^{2}_{3}-\frac{1}{2}(J_{+}J_{-}+J_{-}J_{+})$,  and we may denote the Hamiltonian using $J^{2}$ as $H_{-}=-(J^{2}+\frac{1}{4})$.
Here, the operators act on the physical states which are given by
\begin{equation}\label{36}
  J^{2}|j, \mu \rangle=j(j+1)|j, \mu \rangle
\end{equation}
\begin{equation}\label{37}
  J_{3}|j, \mu \rangle=\mu|j, \mu \rangle
\end{equation}
\begin{equation}\label{38}
  J_{\pm} |j, \mu \rangle=\sqrt{-(j \mp \mu)(j \pm \mu +1)}|j, \mu\pm 1 \rangle.
\end{equation}
And, $|j, \mu \rangle$ can be written in the function space as
\begin{equation}\label{39}
  |j, \mu \rangle=\psi^{(2)}_{(-),j\mu}(z)e^{i\mu \phi}.
  \end{equation}
If we express $J^{2}$ using the operators as \cite{yad}
\begin{equation}\label{40}
\begin{split}
  J^{2}=-\frac{d^{2}}{dz^{2}}+(1-F(z)^{2})(J^{2}_{3}-1/4)-2G'(z)(J_3)-G(z)^{2}-\frac{1}{4}-(U_{1}^{2}(z)+((J_3-1/2)F(z)-\\ G(z))U_{1}(z)+
    U_{1}(z)((J_3-1/2)F(z)-G(z))-\frac{d}{dz}U_{1}(z)\\
    =-\frac{d^{2}}{dz^{2}}+V^{(2)}_{(-)}(z).
  \end{split}
\end{equation}
Choosing our functions $F(z)$ and $G(z)$ as
\begin{equation}\label{41}
  F(z)=\tan h_{q}z,~~~~G(z)=i B_{1}\sec h_{q}z,
\end{equation}
and using a suggestion for each   $U_{1}(z), U_2(z)$ which are given by
\begin{equation}\label{42}
  U_{1}(z)=i \frac{S_1 \cosh_{q}z}{r_1+r_2 \sinh_{q} z} ,~~~~U_{2}(z)=\frac{S_2 \cosh_{q}z}{r_1+r_2 \sinh_{q}z}
   \end{equation}
one can obtain the  potential which is an element of $J^{2}$. Here, $S_1$ and $ S_2$ are constants. Using (\ref{41}) and $U_1(z)$, we obtain
\begin{equation}\label{43}
\begin{split}
 V^{(2)}_{(-)}(z)=q(1/4-\mu)-2i\mu B_1 \tanh_{q}z \sec h_{q} z-B^{2}_{1}\sec h^{2}_{q}z+\\
  \frac{1}{(r_1+r_2 \sinh_{q}z)^{2}} (-2S_{1}(B_1 r_1+q r_2(-1+\mu))+S_1(iS_1+r_2(2\mu-1))\cosh^{2}_{q}z-2S_1(r_1+B_1 r_2-r_1 \mu)\sinh_{q}z).
  \end{split}
\end{equation}
Hence, the parameter restrictions for (\ref{43}) are given by
\begin{equation}\label{44}
  S_1=ir_2 (2\mu-1),~~~~q=-\frac{r^{2}_{1}}{r^{2}_{2}},~~~~B=-\frac{r_1 (1-\mu)}{r_{2}}.
\end{equation}
Using (\ref{44}) in (\ref{43}), we get
 \begin{equation}\label{45}
   V^{(2)}_{(-)}(z)=q(1/4-\mu)+2i\mu(1-\mu) \frac{r_1}{r_2}\tanh_{q}z \sec h_{q} z-\frac{r^{2}_{1}}{r^{2}_{2}}(1-\mu)^{2}\sec^{2} h_{q} z,
 \end{equation}
 where
\begin{eqnarray}\label{45}
  q &=& \frac{1}{2k^{2}K^{2}_{1}}(-2k^{2}K_1-2e^{2}K^{2}_{2}+\frac{r_1(r_1-eK_2(1+2kK_2)r_2)\mp \frac{r_1\sqrt{r_1(r_1-eK_2(1+2kK_2)r_2)}}{r^{2}_{2}}}{r^{2}_{2}}) \\
  \mu &=& \frac{1}{2}\pm \frac{1}{2} \sqrt{1-\frac{2eK_2 r_2}{r_1}-\frac{4eK^{2}_{2}r_2}{r^{2}_{2}}}.\label{47}
\end{eqnarray}
\begin{equation}\label{48}
  E_{n}=\pm \sqrt{q(1/4-\mu)-(\mu+1/2-n)^{2}},~~~~n_{max}< \mu+1/2.
\end{equation}
In order to compare (\ref{48}) with (\ref{14}), one can use $q(1/4-\mu)=1/4\mp \frac{1}{2}(K_2-1/2)^{2}$ and $\lambda_1=\mu+1/2$ and can find $\lambda_2$ and a relation between the parameters $r_1$ and $r_2$.
\section{Conclusions}
In this paper, it is observed that two different superpotentials can generate the same energy spectrum for the corresponding different potential systems.
The component of the vector potential and the Fermi velocity function are chosen as real hyperbolic functions but the potential functions in the Klein-Gordon-like equation are complex functions known as deformed complex  $\mathcal{PT}$ symmetric Scarf II potential whose spectrum is obtained as real. Despite the case in non-relativistic quantum mechanics where the energy is negative, in our system the square of the energy plus a constant term from the potential makes the spectrum as real. We have given a suggestion for a superpotential with a rational term generated two partner potentials which are unsolvable. On the other hand, using the constraints on the potential parameters, one of the partner potential is obtained as deformed complex hyperbolic Scarf II potential while the other partner remained as unsolvable. We have written the well-known solutions of the deformed complex $\mathcal{PT}$ symmetric Scarf II potential and using these solutions the  unsolvable partner potential energy spectrum and spinor solutions are found. Finally, we have found at the operators of $so(2,1)$ Lie algebra in terms of the Dirac system parameters. The energy spectrum results derived from the Casimir operator agree with the results obtained from the Dirac system.

\end{document}